\documentclass[aps,prb,twocolumn,superscriptaddress]{revtex4}
\usepackage{amssymb}
\usepackage{amsmath}
\usepackage{epsfig}
\usepackage{color}
\usepackage{graphics, graphicx}
\usepackage{bbold}
\usepackage{psfrag}
\usepackage{subfigure}
\usepackage{verbatim}
\usepackage{color}
\usepackage[colorlinks,citecolor=blue]{hyperref}

\begin{document}

\title{Quantum spiral spin-tensor magnetism}
\author{Xiaofan Zhou}
\affiliation{Department of Physics, The University of Texas at Dallas, Richardson, Texas
75080, USA}
\affiliation{State Key Laboratory of Quantum Optics and Quantum Optics Devices, Institute
of Laser spectroscopy, Shanxi University, Taiyuan 030006, China}
\author{Xi-Wang Luo}
\affiliation{Department of Physics, The University of Texas at Dallas, Richardson, Texas
75080, USA}
\author{Gang Chen}
\email{chengang971@163.com}
\affiliation{State Key Laboratory of Quantum Optics and Quantum Optics Devices, Institute
of Laser spectroscopy, Shanxi University, Taiyuan 030006, China}
\affiliation{Collaborative Innovation Center of Extreme Optics, Shanxi University,
Taiyuan, Shanxi 030006, China}
\affiliation{Collaborative Innovation Center of Light Manipulations and Applications,
Shandong Normal University, Jinan 250358, China}
\author{Suotang Jia}
\affiliation{State Key Laboratory of Quantum Optics and Quantum Optics Devices, Institute
of Laser spectroscopy, Shanxi University, Taiyuan 030006, China}
\affiliation{Collaborative Innovation Center of Extreme Optics, Shanxi University,
Taiyuan, Shanxi 030006, China}
\author{Chuanwei Zhang}
\email{chuanwei.zhang@utdallas.edu}
\affiliation{Department of Physics, The University of Texas at Dallas, Richardson, Texas
75080, USA}

\begin{abstract}
The characterization of quantum magnetism in a large spin ($\geq 1$) system
naturally involves both spin-vectors and -tensors. While certain types of
spin-vector (e.g., ferromagnetic, spiral) and spin-tensor (e.g., nematic in
frustrated lattices) orders have been investigated separately, the
coexistence and correlation between them have not been well explored. Here
we propose a novel quantum spiral spin-tensor order on a spin-1 Heisenberg
chain subject to a spiral spin-tensor Zeeman field, which can be
experimentally realized using a Raman-dressed cold atom optical lattice. We
develop a method to fully characterize quantum phases of such spiral tensor
magnetism with the coexistence of spin-vector and spin-tensor orders as well
as their correlations using eight geometric parameters. Our method provides
a powerful tool for characterizing spin-1 quantum magnetism and opens an
avenue for exploring novel magnetic orders and spin-tensor
electronics/atomtronics in large-spin systems.
\end{abstract}

\maketitle

\emph{Introduction}.--- Quantum magnetism originates from the exchange
coupling between quantum spins and lies at the heart of many fundamental
phenomena in quantum physics~\cite%
{Auerbach-1994,Schollwock-2008,Sachdev-2008}. In particular, understanding
exotic magnetic orders of strongly correlated quantum spin chains is one
major issue of modern condensed matter physics. Such interacting many-body
systems can give rise to various magnetic orders and the phase transitions
between them~\cite%
{DM-1,DM-2,Manousakis-1991,Haldane-1983,Affleck-1987,AKLT-1987,Islam2013,random2017}%
. Major research efforts have been focused on spin-1/2 systems, where
collinear (e.g., ferromagnetic and antiferromagnetic) and non-collinear
(e.g., spiral) magnetic orders are fully characterized by the spin-vector ($%
\vec{S}$) configurations, including their local orientations and densities
as well as nonlocal correlations~\cite%
{Greiner2011,Esslinger2013,Hulet2015,Bloch2016,Greiner2016,Greiner2017,wu2012,wu2013,wu2017,wu2018,Bloch2019}%
.

Quantum magnetism with large spins, such as spin-1, has also received
considerable attention in recent years, where the large spin could originate
from, for instance, intrinsic orbital degrees of electrons or pseudo-spins
defined by hyperfine states of cold neutral atoms or ions \cite%
{White-1993,Ripoll-2004,Rizzi-2005,Pixley2017,Pixley2018,Natu2015,spin-1SOC2016,Piraud2014,Hamley2012,Ueda2013,SOC-1BEC1,ions1}%
. Mathematically, a full description of a large spin ($\geq 1$) involves not
only rank-1 spin-vectors, but also higher-rank spin-tensors, therefore it is
expected that the resulting quantum magnetism may possess both spin-vector
and tensor orders. So far, spin-vector magnetism of a strongly correlated
spin-1 (or higher) chain has been extensively studied~\cite%
{Lou1999,Lou2005,Hagiwara2005,Manmana2011,Chiara2011,Rodriguez2011,
Weyrauch2017,Pixley2017,Hermele2009,Gorshkov2010,Manmana2011A,yan2013,zhang2014}%
, and the competition between spin interaction and Zeeman field (either
uniform or spiral along the chain) leads to rich phase diagrams.
Certain nematic magnetic orders of spin-tensors (with vanishing spin-vector)
have been investigated in 2-dimensional (D) geometrically frustrated (e.g.,
triangle) lattices \cite%
{triangular1,triangular2,triangular3,triangular4,triangular5,triangular6,triangular7,triangular8,triangular9}%
. However, magnetic orders with the coexistence of these two orders have not
been discovered and a unified description of such magnetic orders is still
lacking. Addressing these two important issues should be of great importance
for the discovery of novel magnetic orders and the exploration of
electronics/atomtronics characterized by spin-tensors.

We restrict to spin-1 magnetic orders, which may be characterized by two
elements: rank-1 spin-vector (represented by an arrow using 3 parameters)
and rank-2 spin-tensor (represented by an ellipsoid using 5 parameters). In
this Letter, we propose a novel type of quantum spiral spin-tensor orders
with the coexistence of spin-vector and tensor orders and develop a
geometric method to describe them. Our main results are:

\textit{i}) We propose an experimental setup for realizing a spin-1
Heisenberg chain subject to a spiral spin-tensor Zeeman field using a Raman
coupled
cold atom optical lattice, which can host the novel quantum spiral
spin-tensor orders.

\textit{ii}) We develop a method for characterizing any spin-1 quantum
magnetism with the order parameters and long-range spin correlations
described by eight geometric parameters originating from spin arrow and
ellipsoid.

\textit{iii}) We obtain the ground-state phase diagram numerically using the
density-matrix renormalization group (DMRG) method and showcases its rich
physics such as a ferromagnetic spiral tensor phase, where the relative
orientation between spin-vector arrow and spin-tensor ellipsoid exhibits
periodic oscillations. In previously studied spiral spin-vector order~\cite%
{Pixley2017}, arrow length, ellipsoid size, and their relative orientations
are uniformly fixed across the lattice chain, while the 2D nematic phase in
a frustrated lattice~\cite{triangular1,triangular2} possesses vanishing
spin-vector arrow and fixed ellipsoid size and directions.

\begin{figure}[t]
\centering
\includegraphics[width = 8.5cm]{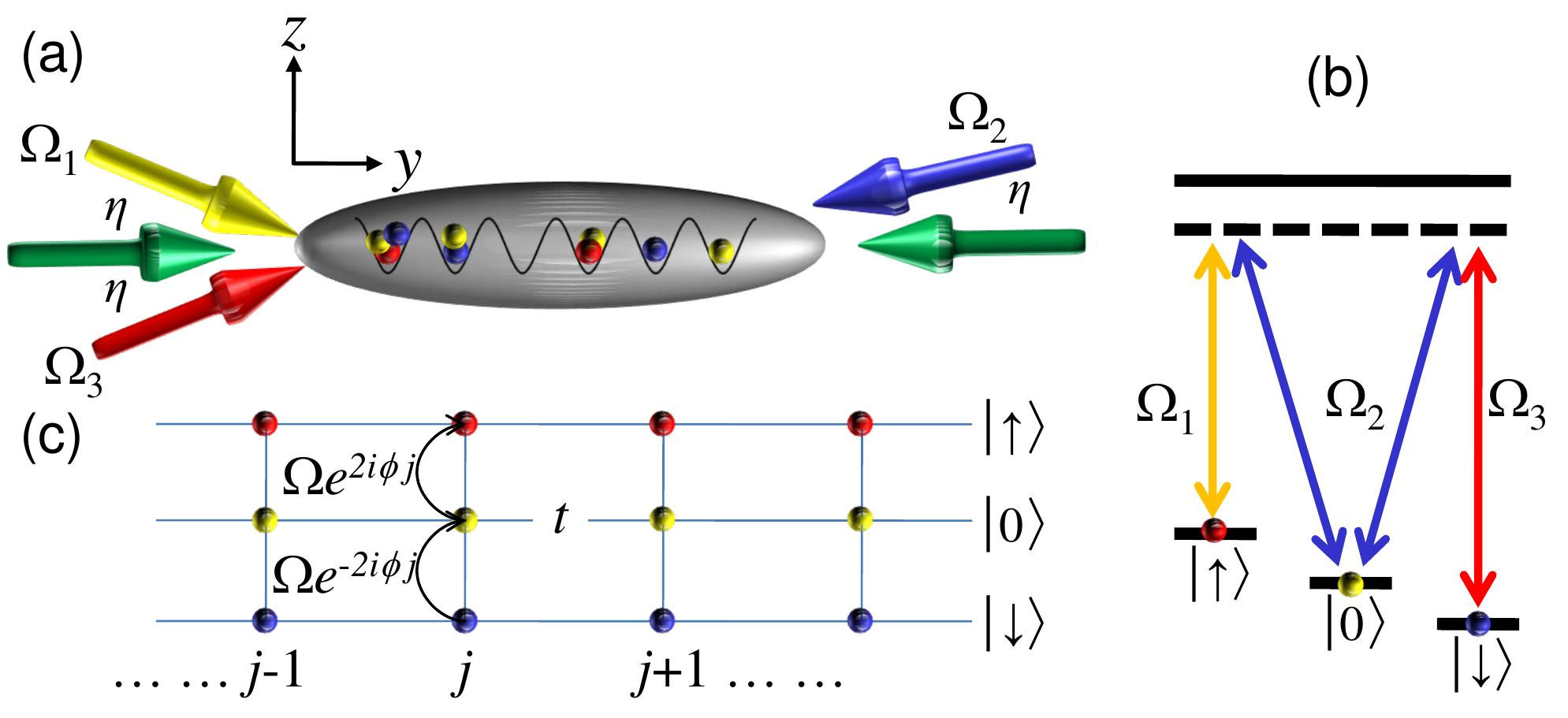} \vskip 0.0cm
\caption{(a) Schematics of the system setup. Green arrows represent the 1D
optical lattice. Yellow, blue and red arrows represent three Raman lasers.
(b) The Raman lasers induce two Raman transitions between spin states $%
\left\vert 0\right\rangle $ and $\!\!\left\vert \uparrow \!\!(\downarrow
)\right\rangle $. (c) Effective description of the model. $t$ is the hopping
between neighboring sites, $\Omega e^{2i\protect\phi j}\left\vert 0\rangle
\!\langle \uparrow \right\vert $ and $\Omega e^{-2i\protect\phi j}\left\vert
\downarrow \rangle \!\langle 0\right\vert $ are the site-dependent couplings
between different spin states.}
\label{experimental}
\end{figure}

\emph{The model}.--- We consider an experimental setup based on ultracold
bosonic atoms in a 1D optical lattice, as shown in Fig.~\ref{experimental}%
(a). Three Raman lasers are used to couple different spin and momentum
states~\cite%
{Ueda2013,Dalibard2011,Goldman2014,spielman2011,spielman2013,Zhai2015,Luo2017}%
,. In particular, a pair of counter-propagating lasers with wavelength $%
\lambda _{\mathrm{L}}$ is used to realize the 1D optical lattice $V_{\mathrm{%
lat}}\left( y\right) =-V_{0}\cos ^{2}\left( k_{\mathrm{L}}y\right) $ along
the $y$-direction, with wavenumber $k_{\mathrm{L}}=2\pi /\lambda _{\mathrm{L}%
}$. All Raman lasers have an angle $\eta $ with respect to the $y$-direction
and they induce two Raman transitions between the spin states $|0\rangle $
and $|\!\!\uparrow \!\!(\downarrow )\rangle $ [as shown in Fig.~\ref%
{experimental}(b)] with the momentum transfer $2k_{\mathrm{R}}$, where $k_{%
\mathrm{R}}=2\pi \cos (\eta )/\lambda _{\Omega }$ with $\lambda _{\Omega }$
the Raman-laser wavelength. Such Raman laser setup induces the
spin-tensor-momentum coupling \cite{Luo2017} for ultra-cold atomic gases,
which has been realized in experiment recently \cite{Li2020}.

The tight-binding Hamiltonian without Raman lasers can be written as $%
H_{0}\!=\!-t\!\sum_{\left\langle i,j\right\rangle ,\sigma }\!{b}_{i\sigma
}^{\dag }{b}_{j\sigma }\!+(U_{0}/2)\sum_{j}{n}_{j}\left( {n}_{j}-1\right)
+(U_{2}/2)\sum_{j}\left( \mathbf{S}_{j}^{2}-2{n}_{j}\right) $, where ${b}%
_{j\sigma }^{\dag }$ (${b}_{j\sigma }$) is the creation (annihilation)
operator with a spin $\sigma $, $n_{j}=\sum_{\sigma }b_{j\sigma }^{\dag
}b_{j\sigma }$ is the density operator, and $\mathbf{S}_{j}=\sum_{\sigma
\acute{\sigma}}{b}_{j\sigma }^{\dag }\mathbf{F}_{\sigma \acute{\sigma}}{b}_{j%
\acute{\sigma}}$ with $\mathbf{F}_{\sigma \acute{\sigma}}$ represent the
total angular momentum $F=1$ spin operators and $\sigma =(\uparrow
,0,\downarrow )$. $t$ is the tunneling amplitude between neighboring sites,
and $U_{0}$ and $U_{2}$ are on-site density and spin interaction strengths.

In the Mott limit of commensurate odd integer filling and $U_{0},U_{2}\gg t$%
, we can get the effective spin Hamiltonian~(see Appendix \ref{Derivation of Hamiltonian}), $H_{\mathrm{spin}%
}^{\mathrm{0}}=\sum_{j}J_{1}\mathbf{S}_{j}\cdot \mathbf{S}_{j+1}+J_{2}\left(
\mathbf{S}_{j}\cdot \mathbf{S}_{j+1}\right) ^{2}$. Typically, $J_{1}<0$ for
repulsive interaction, therefore we parameterize $J_{1}$ and $J_{2}$ on a
unit circle with $J_{1}=\cos \theta $ and $J_{2}=\sin \theta $ and focus on
the parameter region $\theta \in \lbrack 0.5\pi ,1.5\pi ]$. For $J_{2}>J_{1}$
(i.e., $0.5\leq \theta /\pi <1.25$), the system possesses the ferromagnetic
order, which maximizes $\mathbf{S}_{j}\cdot \mathbf{S}_{j+1}$ but minimizes $%
\left( \mathbf{S}_{j}\cdot \mathbf{S}_{j+1}\right) ^{2}$ due to dominating
negative bilinear interaction $J_{1}$. For large negative biquadratic
interaction $J_{2}<J_{1}$ in the region $\left( 1.25<\theta /\pi \leq
1.5\right) $, the ground state should maximize $\left( \mathbf{S}_{j}\cdot
\mathbf{S}_{j+1}\right) ^{2}$ by forming spin singlet between neighboring
sites, which breaks the translational symmetry, leading to the dimer phase~%
\cite{White-1993, Rizzi-2005,Ripoll-2004,Dimer2003}. Typical values for
alkaline atoms are $\theta /\pi =1.26$ for $^{23}$Na, $1.15\pi $ for ${^{7}}$%
Li, $1.242$ for $^{41}$K, and $1.249$ for $^{87}$Rb \cite{Ueda2013}.

The Raman lasers give rise to the site-dependent spin flipping terms $\Omega
e^{2i\phi j}\left\vert 0\rangle \!\langle \uparrow \right\vert $ and $\Omega
e^{-2i\phi j}\left\vert \downarrow \rangle \!\langle 0\right\vert $ [see
Fig.~\ref{experimental}(c)], which would induce a spiral on-site spin-vector
and tensor field. To see this, we write down the tight-binding Hamiltonian
for such Raman processes $H_{\mathrm{\Omega }}\!=\!\!\sqrt{2}\Omega
\sum_{j}(e^{2i\phi j}\hat{b}_{j\uparrow }^{\dag }\hat{b}_{j0}+e^{-2i\phi j}%
\hat{b}_{j0}^{\dag }\hat{b}_{j\downarrow }+\mathrm{H.c.})$, $\Omega $ is the
Raman coupling strength, $\phi =\pi \cos (\eta )\lambda _{\mathrm{L}%
}/\lambda _{\Omega }$ describes the flux and can be tuned by the angle $\eta
$. In the Mott limit, $H_{\mathrm{\Omega }}$ can be treated as spiral
spin-vector and spin-tensor Zeeman fields $H_{\mathrm{spin}}^{\mathrm{\Omega
}}=2\Omega \sum_{j}\left[ \cos \left( 2\phi j\right) S_{j}^{x}-\sin \left(
2\phi j\right) N_{j}^{yz}\right] $, where $N^{\alpha \beta }=\{S^{\alpha
},S^{\beta }\}/2-\delta _{\alpha \beta }\mathbf{S}^{2}/3$ with $\{\}$ the
anticommutation relation and $\alpha (\beta )=(x,y,z)$, $\Omega $ is the
Zeeman field strength, and $\phi $ is the spiral period of the field. The
total Hamiltonian of our system reads%
\begin{equation}
H_{\mathrm{spin}}=H_{\mathrm{spin}}^{\mathrm{0}}+H_{\mathrm{spin}}^{\mathrm{%
\Omega }}.  
\label{HS}
\end{equation}%
The competition between spin interaction $H_{\mathrm{spin}}^{\mathrm{0}}$
and spiral on-site field $H_{\mathrm{spin}}^{\mathrm{\Omega }}$ may induce
many novel spin-tensor magnetic phases, where both spin-vector and
spin-tensor have to be considered to fully describe these quantum magnetic
orders.

\begin{figure}[t]
\centering
\includegraphics[width = 8.5cm]{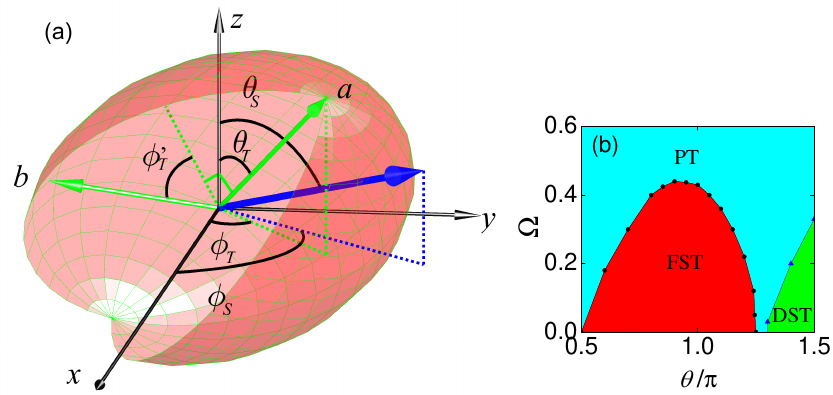} %
\hskip 0.0cm
\caption{(a) The description of a spin-1 magnetic order. The blue arrow
denotes the spin-vector $\vec{S}$, in which $\protect\theta _{S}$ and $%
\protect\phi _{S}$ are polar angle and azimuth angle with length $l_{S}$.
The red ellipsoid is the spin-tensor $T$, in which the green arrows are
ellipsoid's orientations $\vec{v}_{T}^{n}$ ($n=a,b$) with the principle axes
lengths $l_{T}^{n}$. The Euler angles $\protect\theta _{T}$, $\protect\phi %
_{T}$ and $\protect\phi _{T}^{\prime }$ are used to determine the
orientation of ellipsoid, in which $\protect\theta _{T}$ is the polar angle
of $\vec{v}_{T}^{a}$, $\protect\phi _{T}$ is the azimuth angle of $\vec{v}%
_{T}^{a}$, and $\protect\phi _{T}^{,}$ is the angle between $\vec{v}_{T}^{b}$
and the plane formed by $z$-axis and $\vec{v}_{T}^{a}$. (b) The phase
diagram of the Hamiltonian (\protect\ref{HS}) with respect to ($\protect%
\theta ,\Omega $) for $\protect\phi /\protect\pi =1/6$, which includes
paramagnetic tensor (PT), ferromagnetic spiral tensor (FST) and dimer spiral
tensor (DST) phases.}
\label{phase_diagram_theta_Omega_Vectors_Tensor}
\end{figure}

\emph{Description of spin-1 magnetic order}.--- The local magnetic order of
a spin-1 system can be described by the local densities of 8 spin-moments
(i.e., 3 spin-vectors and 5 spin-tensors). Here we consider the local
spin-vector $\vec{S}_{j}=(\langle S_{j}^{x}\rangle ,\langle S_{j}^{y}\rangle
,\langle S_{j}^{z}\rangle )^{T}$ and the spin-tensor fluctuation matrix $%
T_{j}$ whose elements are tensor moments $T_{j}^{\alpha \beta }=\langle
\{S_{j}^{\alpha },S_{j}^{\beta }\}\rangle /2-\langle S_{j}^{\alpha }\rangle
\langle S_{j}^{\beta }\rangle $. Geometrically, $\vec{S}_{j}$ is
characterized by an arrow and $T_{j}$ by an ellipsoid (with principle axis
lengths $l_{T}^{n}\left( j\right) $ ($n=a,b,c$) and orientations $\vec{v}%
_{T}^{n}\left( j\right) $ given by the square-roots of the eigenvalues and
eigenvectors of $T_{j}^{\alpha \beta }$~\cite{Bharath}). To quantitatively
characterize and geometrically visualize the magnetic order, we choose 8
independent geometric parameters that could fully describe the spin arrow
and ellipsoid: the length $l_{S}$ and spherical coordinates $\theta _{S}$, $%
\phi _{S}$ of the arrow, the two axis lengths $l_{T}^{a,b}$ with the third
axis length $l_{T}^{c}= \sqrt{{2-(l_S)}^2- {(l_{T}^{a})}^2 -{(l_{T}^{b})}^2}$
and orientational Euler angles $\theta _{T}$, $\phi _{T}$, $\phi
_{T}^{\prime }$ of the ellipsoid, as shown in Fig.~\ref%
{phase_diagram_theta_Omega_Vectors_Tensor}(a).

Beside the local densities, the characterization of long-range correlations
of the magnetic order requires spin-vector and spin-tensor correlations. In
our Heisenberg model, only the same-spin-moment correlations are relevant
because of the same-spin-moment interactions, and correlation function of
spin moment $\hat{O}$ over a distance $r$ can be defined as $\mathbb{C}(\hat{%
O},r)=(1/L)\sum_{j}[\langle \hat{O}_{j}\hat{O}_{j+r}\rangle _{\nu }-\langle
\hat{O}_{j}\rangle _{\nu }\langle \hat{O}_{j+r}\rangle _{\nu }]$, with $%
\langle ~\rangle _{\nu }$ the average value of the $\nu $-fold degenerate
ground-states. The correlations can have the same geometrical representation
as the local densities. The spin-vector correlation is described by an arrow
$\vec{\mathbb{S}}_{r}$ with $\mathbb{S}_{r}^{\alpha }=\mathbb{C}\left(
S^{\alpha },r\right) $ and spin arrow lengths $S_{r}$, while the spin-tensor
correlation $\mathbb{T}_{r}$ is described by an ellipsoid with the principle
axis lengths $l_{r}^{n}$ given by the tensor-correlation matrix $\mathbb{T}%
_{r}^{\alpha \beta }=\mathbb{C}(N^{\alpha \beta },r)$.

\emph{Spiral spin-tensor magnetism}.--- The numerical ground-state phase
diagram of the Hamiltonian~(\ref{HS}) can be obtained through the DMRG
calculation \cite{DMRG-1,DMRG-2}, where the length of the spin chain $L$ is
up to 96 sites, and we keep the maximum states at 200 and achieve truncation
errors of $10^{-8}$. The resulting phase diagram for $\phi /\pi =1/6$ is
plotted in the $\theta -\Omega $ plane in Fig.~\ref%
{phase_diagram_theta_Omega_Vectors_Tensor}(b). There are three different
phases: the ferromagnetic spiral tensor phase (FST) for $0.5\leq \theta /\pi
<1.25$ and the dimer spiral tensor phase (DST) for $1.25<\theta /\pi \leq
1.5 $, both in the small spiral on-site field $\Omega $ region, and the
paramagnetic tensor phase (PT) for the large $\Omega $. All these phases
possess both spiral spin-vector and -tensor densities and can be
distinguished by spin correlations: \textit{i}) For the FST phase, there are
spin-vector and spin-tensor long-range correlations with nonzero
ferromagnetic order $c_{F}\equiv \langle S_{1}^{z}S_{L/2}^{z}\rangle $;
\textit{ii}) For the DST phase, the dimer-vector and dimer-tensor
correlations are long-range with nonzero dimer order $c_{D}\equiv \langle
D_{1}^{S^{z}}\!D_{L/2}^{S^{z}}\rangle $ (see Appendix \ref{Dimer spiral tensor phase});
\textit{iii}) For the
PT phase, there is no any long-range correlations. For other spiral period $%
\phi $, quantum phase diagrams are similar, except that the phase
transitions may occur at different critical points and the period of the
spiral modulation varies in the same way as the spiral period $\phi $.

\begin{figure}[t]
\centering
\includegraphics[width = 8.9cm]{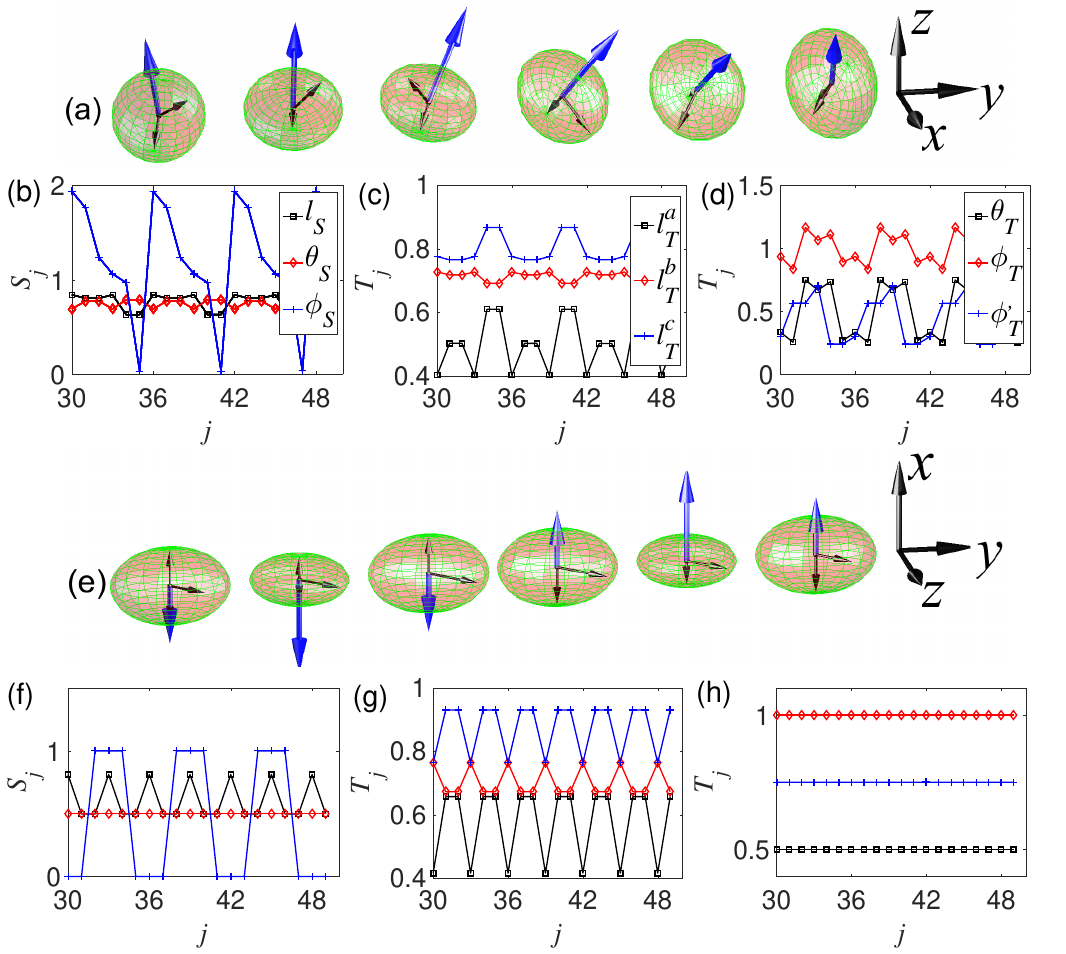} \hskip 0.0cm
\caption{(a)(e) Schematic diagram of spin-vector density arrows $\vec{S_{j}}$
and spin-tensor density ellipsoids $T_{j}$. (b)(f) Spatial distributions of
the spin-vector density arrows $\vec{S_{j}}$. (c,d,g,h) Spatial
distributions of the spin-tensor density ellipsoids $T_{j}$. (a)-(d) FST
phase with $\protect\theta /\protect\pi =0.9$ and $\Omega =0.4$. (e)-(h) PT
phase with $\protect\theta /\protect\pi =0.9$ and $\Omega =0.5$. $\protect%
\phi /\protect\pi =1/6$ for all subfigures. }
\label{FST_PST_angle_length}
\end{figure}

First we consider the region $0.5\pi \leq \theta <1.25\pi $ with the weak $%
\Omega $ (the bottom-left part in the phase diagram), where the spin
interactions are still dominant and the ferromagnetic order remains. At $%
\Omega =0$, the spin-vector arrow points to certain direction (assumed to be
the $z$ axis) that does not change along the chain, and the spin-tensor
ellipsoid is a flat disk in the $x$-$y$ plane [i.e., $l_{T}^{a}=0$, $\theta
_{T}=0$ (i.e., $a$ axis is parallel to $z$ axis)]~(see Appendix \ref{Spin-vector and spin-tensor magnetic orders}). In the
presence of the spiral Zeeman field ($\Omega >0$), the spin-vector density
arrows $\vec{S_{j}}$, the spin-tensor density ellipsoids $T_{j}$, and their
relative orientations become oscillating periodically along the chain,
forming spiral loops in the Bloch sphere (see Appendix \ref{Spin-vector and spin-tensor magnetic orders}) and leading to the
FST phase, where the local spiral magnetism and long-range correlations
coexist. Fig.~\ref{FST_PST_angle_length}(a) shows the arrows $\vec{S}_{j}$
and ellipsoids $T_{j}$ in one spatial period, which possess spiral structure.

For the quantitative characterization of the spiral tensor order, we plot
the spatial distributions of $\left[ l_{S}(j) ,\theta _{S}(j) ,\phi _{S}(j) %
\right] $ for the vector-density arrows, the principle axis lengths $\left[
l_{T}^{a}(j) ,l_{T}^{b}(j) ,l_{T}^{c}(j) \right] $, and the orientational
Euler angles $\left[ \theta _{T}(j) ,\phi _{T}(j) ,\phi _{T}^{,}(j) \right] $
of the ellipsoids in Figs.~\ref{FST_PST_angle_length}(b)-\ref%
{FST_PST_angle_length}(d). We find $l_{S}(j)$, $\theta _{S}(j)$ and $%
\phi_{S}(j)$ oscillate along the chain with the period same as the spiral
field, and $\phi_{S}(j)$ changes from $0$ to $2\pi $ during one period [see
Fig.~\ref{FST_PST_angle_length}(b)]. Correspondingly, the arrows $\vec{S}_{j}
$ within one period form a circular loop around the $z$ axis (see Appendix \ref{Spin-vector and spin-tensor magnetic orders}).
For the spin-tensor density ellipsoids $T_{j}$, beside the modulation in its
size $l_{T}^{n}(j)$ [see Fig.~\ref{FST_PST_angle_length}(c)], the
corresponding axes form a twisted loop ($8$-shaped)~(see Appendix \ref{Spin-vector and spin-tensor magnetic orders}), leading
to relative rotations between spin-vector density arrows and spin-tensor
density ellipsoids [$\theta _{T}(j)$, $\phi _{T}(j)$ and $\phi _{T}^{\prime
}(j)$ oscillate differently from $\theta _{S}(j)$, $\phi _{S}(j)$], as shown
in Fig.~\ref{FST_PST_angle_length}(d). Such spiral magnetic configuration
originates from the competition between the on-site spin-vector potential [$%
\cos \left( 2\phi j\right) S_{j}^{x}$] and spin-tensor potential [$-\sin
\left( 2\phi j\right) N_{j}^{yz}$]. Without the spin-tensor field (i.e. only
spin-vector field), the spin-vector density arrows and spin-tensor density
ellipsoids would rotate similarly with fixed relative orientation [i.e., $%
\theta _{T}(j)-\theta _{S}(j)$, $\phi _{T}(j)-\phi _{S}(j)$ and $\phi
_{T}^{\prime }(j)$ uniform along the chain], and all ellipsoids would have a
fixed size [i.e., $l_{T}^{n}(j)$ uniform along the chain]~(see Appendix \ref{The results due to the spin-vector potential}).
Moreover, the ferromagnetic spiral order is characterized by the long-range
correlation of both spin-vector $\vec{\mathbb{S}}_{r}$ and spin-tensor $%
\mathbb{T}_{r}$~(see Appendix \ref{Spin-vector and spin-tensor magnetic orders}).

\begin{figure}[t]
\includegraphics[width = 8.7cm]{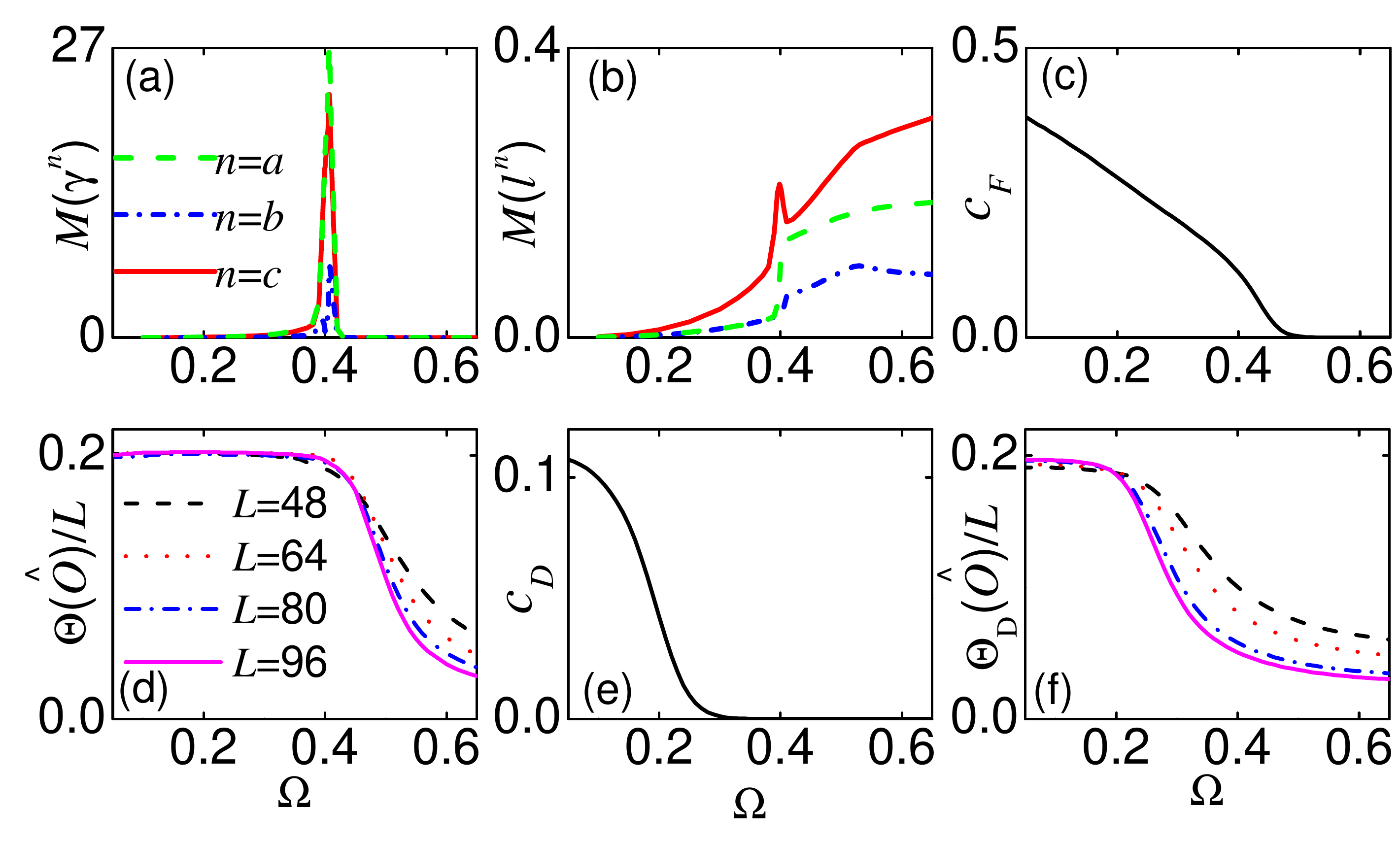} \hskip 0.0cm \centering
\caption{(a)-(c) Oscillation amplitude of angles $M(\protect\gamma ^{n})$,
lengths $M(l^{n})$ and ferromagnetic order $c_{F}$ as a function of $\Omega $%
. (d) The correlation lengths $\Theta (\hat{O})/L$ as functions of $\Omega $
for different lattice length $L$. (e) Dimer order $c_{D}$ as a function of $%
\Omega $. (f) Dimer correlation lengths $\Theta _{\mathrm{D}}(\hat{O})/L$ as
functions of $\Omega $ for different $L$, with $\protect\theta /\protect\pi %
=1.4$. In (d)(f), we only plot $\hat{O}=S^{z}$ as an example, others are
shown in Supplemental Material~\protect(see Appendix \ref{Correlation lengths near phase transitions}).
$\protect\theta /%
\protect\pi =0.9$ in (a)-(d), $L=96$ in (a)-(c) and (e). Common parameter $%
\protect\phi /\protect\pi =1/6$. }
\label{phase transition Omega}
\end{figure}

As we increase $\Omega $, the spin-vector rotation loop in a period first
enlarges and then shrinks to a narrow ellipse (cigar-shaped). For a strong $%
\Omega $, the system undergoes a second-order phase transition from the FST
phase to the PT phase, where the long-range correlations and ferromagnetic
order $c_{F}$ vanish [see Fig.~\ref{phase transition Omega}(c)]. In the PT
phase, the on-site Zeeman field in Hamiltonian~(\ref{HS}) dominates, and all
spin-vector density arrows are parallel to the Zeeman field [$\cos \left(
2\phi j\right) S_{j}^{x}$] with length modulations. The corresponding local
magnetic densities are shown in Fig.~\ref{FST_PST_angle_length}(e). Clearly,
the spin-vector density loop shrinks into a line on the $x$-axis in the PT
phase ($\theta _{S}=\pi /2$ and $\phi _{S}=0,\pi $).

The rotation loop of the tensor ellipsoids changes similarly as we increase $%
\Omega $: the loop first enlarges then shrinks into a line after the phase
transition, where only the sizes $l_{T}^{n}(j)$ (not the direction) of the
ellipsoids oscillate (see Appendix \ref{Spin-vector and spin-tensor magnetic orders}), as shown in Figs.~\ref%
{FST_PST_angle_length}(g) and \ref{FST_PST_angle_length}(h). The oscillation
of $l_{T}^{n}(j)$ makes the PT phase different from regular paramagnetic
phase. In the PT phase, the vector correlation arrows $\vec{\mathbb{S}}_{r}$
and tensor correlation ellipsoids $\mathbb{T}_{r}$ decay exponentially (in
size) with the distance $r$, indicating that there does not exist any
long-range order (see Appendix \ref{Spin-vector and spin-tensor magnetic orders}).

Now we turn to the parameter region $1.25<\theta /\pi \leq 1.5$. The system
still stays in the PT phase for a large $\Omega $. For a small $\Omega $,
the negative biquadratic interaction dominates, leading to the second-order
phase transition to the DST phase. In this phase, the local spin-vector and
spin-tensor densities behave similarly as in the PT phase, forming lines
instead of spiral loops~(see Appendix \ref{Spin-vector and spin-tensor magnetic orders}). Interestingly, there exist spiral
loops for the dimer-vector densities $\mathrm{D}^{\vec{S}}_{j}$ and
dimer-tensor densities $\mathrm{D}^T_{j}$~(see Appendix \ref{Dimer spiral tensor phase}), thanks to the
presence of the on-site spin-tensor field. The ordinary correlations of both
spin-vector $\vec{\mathbb{S}}_{r}$ and spin-tensor $\mathbb{T}_{r}$ decay
similarly as the PT phase~(see Appendix \ref{Dimer spiral tensor phase}). However, there exist long-range
correlations for both dimer-vector $\mathbb{D}^{\vec{S}}_{r}$ and
dimer-tensor $\mathbb{D}^T_{r}$~(see Appendix \ref{Dimer spiral tensor phase}), and the dimer order $c_{D}$
emerges.

\emph{Phase transitions}.--- The phase transitions between above phases can
be characterized by the critical behaviors of the local densities and the
correlations of the spin-vectors and -tensors. When $\Omega $ increases from
zero, the spiral loops of spin-vector density arrows $\vec{S}_{j}$ and
spin-tensor density ellipsoids' axes emerge from initial uniform
distribution, become larger, then shrink, and finally disappear~at the phase
transition (see Appendix \ref{Spin-vector and spin-tensor magnetic orders}). The spin-tensor density ellipsoids $T_{j}$ have
two critical behaviors when crossing the phase transition: \textit{i}) the
angles $\gamma _{j}^{n}=\vec{S}_{j}\angle \vec{v}_{T}^{n}(j)$ shows a large
oscillation in the real space, which can be characterized by a sharp peak
for the oscillatory amplitude of angles $M(\gamma ^{n})=\max (\gamma
_{j}^{n})-\min (\gamma _{j}^{n})$ located at the phase boundary, as clearly
shown in Fig.~\ref{phase transition Omega}(a); \textit{ii}) the oscillatory
amplitude of ellipsoid's axis lengths $M(l_{T}^{n})=\max [l_{T}^{n}\left(
j\right) ]-\min [l_{T}^{n}\left( j\right) ]$ have sharp features at the
critical point, as shown in Fig.~\ref{phase transition Omega}(b). In
addition, the ferromagnetic order $c_{F}$ decreases and vanishes across the
phase transition, as shown in Fig.~\ref{phase transition Omega}(c).

The transition from the PT to FST or DST phases corresponds to the formation
of the long-range order, therefore the transition should also be captured by
more essential correlation lengths. The numerical results for the
spin-vector (-tensor) correlation length $\Theta (\hat{O})=\sqrt{\frac{%
\sum_{j\neq L/2}(j-L/2)^{2}\langle \hat{O}_{j}\hat{O}_{L/2}\rangle }{%
2\sum_{j\neq L/2}\langle \hat{O}_{j}\hat{O}_{L/2}\rangle }}$~\cite%
{Campostrini2014} ($\hat{O}=S^{y,z},T^{yy,zz,xy,xz}$) for the transition
between PT and FST phases are shown in Fig.~\ref{phase transition Omega}(d),
which show that there are (no) spin-vector and -tensor long-range
correlations for the FST (PT) phase. The critical point $\Omega _{c}$ in the
thermodynamic limit is determined by the crossing point between $\Theta (%
\hat{O})/L$ curves for different finite lattice lengths, as shown in Fig.~%
\ref{phase transition Omega}(d).

Similarly, Figs.~\ref{phase transition Omega}(e) and \ref{phase transition
Omega}(f) show the dimer order $c_{D}$ and dimer-vector(-tensor) correlation
length $\Theta _{\mathrm{D}}(\hat{O})/L$ for finite lattice lengths, where $%
\Theta _{\mathrm{D}}(\hat{O})=\Theta (D_{\hat{O}})$ ($\hat{O}={%
S^{x,y,z},T^{xx,yy,zz,xy,xz,yz}}$) is used to determine the critical point
between PT and DST phases. The dimer correlation length of a finite lattice
length retains and then decays. In the decaying region, the dimer
correlation lengths of several finite lattices cross at one point. In the
process of increasing $\Omega $, the dimer-vector(-tensor) density spiral
loop emerges from a uniform distribution, becomes larger, then shrinks, and
finally disappears (see Appendix \ref{Dimer spiral tensor phase}).

\emph{Discussion and conclusions}.--- The phase transition is related with
the symmetry of the Hamiltonian. For instance, in the FST phase, the ground
state is 4-fold degenerate due to spontaneously symmetry breaking of the $%
Z_{2}$ exchange symmetry $S^{y}\leftrightarrow S^{z}$ and $Z_{2}$ reflection
symmetry $S^{y(z)}\leftrightarrow -S^{y(z)}$ of the Hamiltonian (\ref{HS}),
leading to nonzero $\left\langle S^{z}\right\rangle $ and $\left\langle
S^{y}\right\rangle $. However, in the PT phase, the ground state is
non-degenerate, yielding $\left\langle S^{z}\right\rangle =0$ and $%
\left\langle S^{y}\right\rangle =0$, thus the spiral loop shrinks into a
line ($l_{S}$ oscillate, $\theta _{S}=\pi /2$, $\phi _{S}=0,\pi $) [see Fig.~%
\ref{FST_PST_angle_length}(f)].

Finally we remark that the local spin states are mixed states for the
strongly correlated spin chain, therefore another useful representation of
higher-spin pure states, Majorana stars~\cite%
{Ueda2013,Majorana1932,Majorana1998,Majorana2012,Majorana2013zhai,Majorana2014}%
, is not suitable for studying correlated quantum magnetism.
The spin-tensor can appear not only as an on-site Zeeman field, but also as
interactions between nearest-neighbor sites. We show the magnetic order with spin-tensor interaction in Appendix \ref{Magnetic order with spin-tensor interaction}.

In summary, we propose a new type of quantum magnetism, the spiral
spin-tensor order, in a spin-1 Heisenberg chain subject to a spiral
spin-tensor Zeeman field. We characterize such quantum spiral spin-tensor
orders and their phase transitions using local spin-vector and -tensor
densities and their correlations, which can be visualized using eight
geometric parameters. To detect such magnetic orders, we can isolate the
sites of interest using additional site-resolved potentials and measure
their local spin states and non-local spin correlations~\cite%
{Greiner2011,Greiner2016,Greiner2017,Bloch2016}. Our method can also be used
to describe spin-1 quantum magnetism formed by trapped ion array with
tunable long-range interactions \cite{ions1}. Our work should lay the
foundation for exploring strongly-correlated quantum magnetism in a large
spin system and pave the way for engineering novel types of spin-tensor
electronic/atomtronic devices.

\begin{acknowledgments}
\textbf{Acknowledgements}: X. Z., G. C., and S. J. are supported by National
Key R\&D Program of China under Grants No.~2017YFA0304203; the NSFC under
Grants No.~11674200 and No.~11804204; and 1331KYC. X. L. and C. Z. are
supported by AFOSR (FA9550-16-1-0387), ARO (W911NF-17-1-0128), and NSF
(PHY-1806227),
\end{acknowledgments}

\appendix

\section{\textbf{Derivation of Hamiltonian~(\ref{HS})}}
\label{Derivation of Hamiltonian}

The tight-binding Hamiltonian~(\ref{HS}) can be written as three terms
\begin{equation}
H=H_{\mathrm{t}}+H_{\Omega }+H_{\mathrm{U}},
\end{equation}%
where
\begin{eqnarray}
H_{\mathrm{t}} &=&-t\sum_{\left\langle i,j\right\rangle ,\sigma }\hat{b}%
_{i\sigma }^{\dag }\hat{b}_{j\sigma }, \\
H_{\Omega } &=&\!2\Omega \cos \left( 2\phi j\right) S_{j}^{x}\!-\!2\Omega
\sin \left( 2\phi j\right) N_{j}^{yz}, \\
H_{\mathrm{U}} &=&\frac{U_{0}}{2}\sum_{j}\hat{n}_{j}\left( \hat{n}%
_{j}-1\right) +\frac{U_{2}}{2}\sum_{j}\left( \mathbf{S}_{j}^{2}-2\hat{n}%
_{j}\right) .
\end{eqnarray}%
The effective spin model is obtained through the projection
\begin{equation}
H_{\mathrm{spin}}=P_{s}H^{\prime }P_{s},
\end{equation}%
where $P_{s}$ is the projection operator that projects the states into the
low-energy subspace with a filling $N$ ($N$ is an odd integer), $H^{\prime }$
is the Hamiltonian after the Schrieffer-Wolff transformation~\cite%
{Schrieffer-Wolff}
\begin{eqnarray}
H^{\prime }\!\!\!\! &=&\!\!e^{-iO}He^{iO}  \notag \\
&=&\!\!H-i\left[ O,H\right] +\frac{1}{2!}\left[ O,\left[ O,H\right] \right]
+\cdots  \notag \\
&=&\!\!H_{t}+H_{\mathrm{U}}+H_{\Omega }-i\left( \left[ O,H_{\mathrm{t}}%
\right] +\left[ O,H_{\mathrm{U}}\right] +\left[ O,H_{\Omega }\right] \right) \notag \\
&+& \frac{1}{2!}\left( \left[ O,\left[ O,H_{\mathrm{t}}\right] \right] +\left[
O,\left[ O,H_{\mathrm{U}}\right] \right] +\left[ O,\left[ O,H_{\Omega }%
\right] \right] \right) \notag \\
&+& \cdots .
\end{eqnarray}

\begin{figure*}[t]
\centering
\includegraphics[width = 18cm]{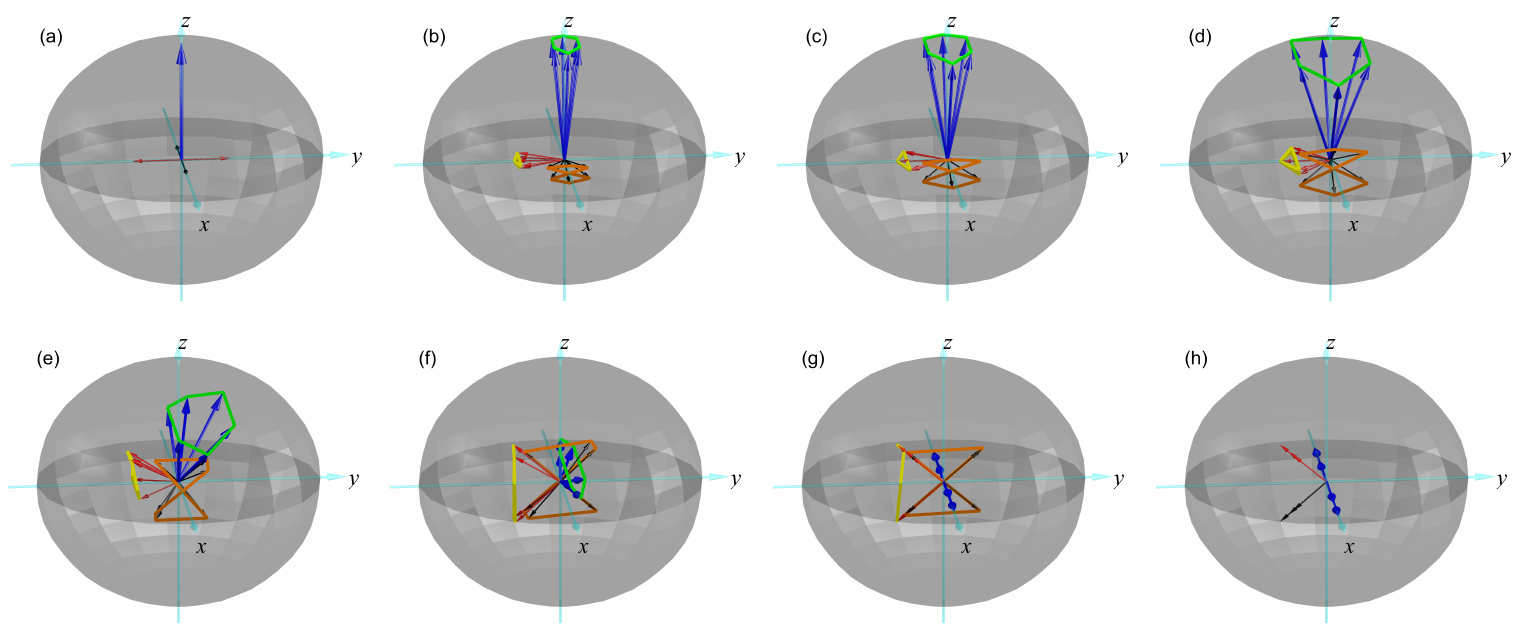} \hskip 0.0cm
\caption{Spiral loops for the phase transition F-FST-PT. Blue arrows
represent spin-vectors, and green line represents the corresponding loop for
the spiral loop of the spin-vector density arrows $\vec{S_{j}}$. Red and
black arrows represent two axes of ellipsoids, while yellow and orange lines
represent corresponding loops for the spin-tensors density ellipsoids $T_{j}$%
. (a) $\Omega =0.0$, (b) $\Omega =0.1$, (c) $\Omega =0.2$, (d) $\Omega =0.38$%
, (e) $\Omega =0.41$, (f) $\Omega =0.42$, (g) $\Omega =0.45$, (h) $\Omega
=0.5$. The phase transition occurs at $\Omega _{c}=0.425$. In all
subfigures, $\protect\theta /\protect\pi =0.9$, $\protect\phi /\protect\pi %
=1/6$ and $L=96 $.}
\label{loop_FST}
\end{figure*}

In the Mott insulator region, the hopping term $H_{\mathrm{t}}$ is small. We
choose $O$ such that $H_{\mathrm{t}}-i\left[ O,H_{\mathrm{U}}\right] =0$,
thus
\begin{eqnarray}
H^{\prime } &=& H_{\mathrm{U}}+H_{\Omega }-\frac{i}{2}\left[ O,H_{\mathrm{t}}%
\right] -i\left[ O,H_{\Omega }\right] \notag \\
&+& \frac{1}{2!} ( \left[ O,\left[
O,H_{\mathrm{t}}\right] \right] + \left[ O,\left[ O,H_{\Omega }\right] \right]) +\cdots .
\end{eqnarray}
Up to the second order of $O$ ($H_{\mathrm{t}}$ is the first order), we have
\begin{equation}
H^{\prime }\!=\!H_{\mathrm{U}}+H_{\Omega }\!-\!\frac{i}{2}\left[ O,H_{%
\mathrm{t}}\right] \!-\!i\left[ O,H_{\Omega }\right] \!+\!\frac{1}{2}\left[
O,\left[ O,H_{\Omega }\right] \right] .
\end{equation}

Denote $P_{d}$ as the projection operator that projects the states into the
subspace of the high-energy with $N+1$ filling. Because $H_{\mathrm{t}%
}=P_{s}HP_{d}+P_{d}HP_{s}=P_{s}H_{\mathrm{t}}P_{d}+P_{d}H_{\mathrm{t}}P_{s}$%
, we have
\begin{eqnarray}
O\!\!\! &=&\!\!P_{s}OP_{d}+P_{d}OP_{s}  \notag \\
&=&\!\!\!\!-i\frac{P_{s}H_{\mathrm{t}}P_{d}}{P_{d}H_{\mathrm{U}%
}P_{d} \!\! - \!\! P_{s}H_{\mathrm{U}}P_{s}}
+i\frac{P_{d}H_{\mathrm{t}}P_{s}}{P_{d}H_{%
\mathrm{U}}P_{d} \!\!-\!\! P_{s}H_{\mathrm{U}}P_{s}},
\end{eqnarray}
where, $P_{d}H_{\mathrm{U}}P_{d}-P_{s}H_{\mathrm{U}}P_{s}\approx \langle
P_{d}H_{\mathrm{U}}P_{d}\rangle -\langle P_{s}H_{\mathrm{U}}P_{s}\rangle $. $%
O$ is not affected by $H_{\Omega }$.
Thus the effective spin Hamiltonian
\begin{eqnarray}
H_{\mathrm{spin}} \!\!&=& \!\! P_{s}H_{\mathrm{U}}P_{s}+P_{s}H_{\Omega }P_{s}-\frac{i}{2}%
P_{s}\left[ O,H_{t}\right] P_{s} \notag \\
&-&iP_{s}\left[ O,H_{\Omega }\right] P_{s} +\frac{1}{2}P_{s}\left[ O,\left[ O,H_{\Omega }\right] \right] P_{s},
\end{eqnarray}%
where, $P_{s}H_{\mathrm{U}}P_{s}=\mathrm{const}$, $P_{s}H_{\Omega
}P_{s}=2\Omega \cos \left( 2\phi j\right) S_{j}^{x}-2\Omega \sin \left(
2\phi j\right) N_{j}^{yz}$,
\begin{eqnarray*}
-\frac{i}{2}P_{s}\left[ O,H_{\mathrm{t}}\right] P_{s} &=&-\frac{1}{2}\frac{%
P_{s}H_{\mathrm{t}}P_{d}H_{\mathrm{t}}P_{s}}{P_{d}H_{\mathrm{U}%
}P_{d}-P_{s}H_{\mathrm{U}}P_{s}} \\
&=&\sum_{j}J_{1}\mathbf{S}_{j}\cdot \mathbf{S}_{j+1}+J_{2}\left( \mathbf{S}%
_{j}\cdot \mathbf{S}_{j+1}\right) ^{2}.
\end{eqnarray*}%
Here $J_{1}$ and $J_{2}$ are given by~\cite{Imambekov-2003}
\begin{eqnarray*}
-\frac{J_{1}}{t^{2}} &=&\frac{2\left( 15+20n+8n^{2}\right) }{15\left(
U_{0}+U_{2}\right) }-\frac{16\left( 5+2n\right) n}{75\left(
U_{0}+4U_{2}\right) }, \\
-\frac{J_{2}}{t^{2}} &=&\frac{2\left( 15+20n+8n^{2}\right) }{45\left(
U_{0}+U_{2}\right) }+\frac{4\left( 1+n\right) \left( 3+2n\right) }{9\left(
U_{0}-2U_{2}\right) } \notag \\
&+&\frac{4n\left( 5+2n\right) }{225\left(
U_{0}+4U_{2}\right) },
\end{eqnarray*}%
$-iP_{s}\left[ O,H_{\Omega }\right] P_{s}=0$, and $P_{s}\left[ O,\left[
O,H_{\Omega }\right] \right] P_{s}\sim t^{2}\Omega /U^{2}\ll t^{2}/U$. After
ignoring the constant term, the final effective spin Hamiltonian becomes,
\begin{eqnarray}
H_{\mathrm{spin}} &=& \sum_{j}J_{1}\mathbf{S}_{j}\cdot \mathbf{S}%
_{j+1}+J_{2}\left( \mathbf{S}_{j}\cdot \mathbf{S}_{j+1}\right)^{2} \notag \\
&+& 2\Omega
\cos \left( 2\phi j\right) S_{j}^{x}-2\Omega \sin \left( 2\phi j\right)
N_{j}^{yz}.
\end{eqnarray}

\section{\textbf{Spin-vector and spin-tensor magnetic orders}}
\label{Spin-vector and spin-tensor magnetic orders}

\begin{figure*}[t]
\includegraphics[width = 17cm]{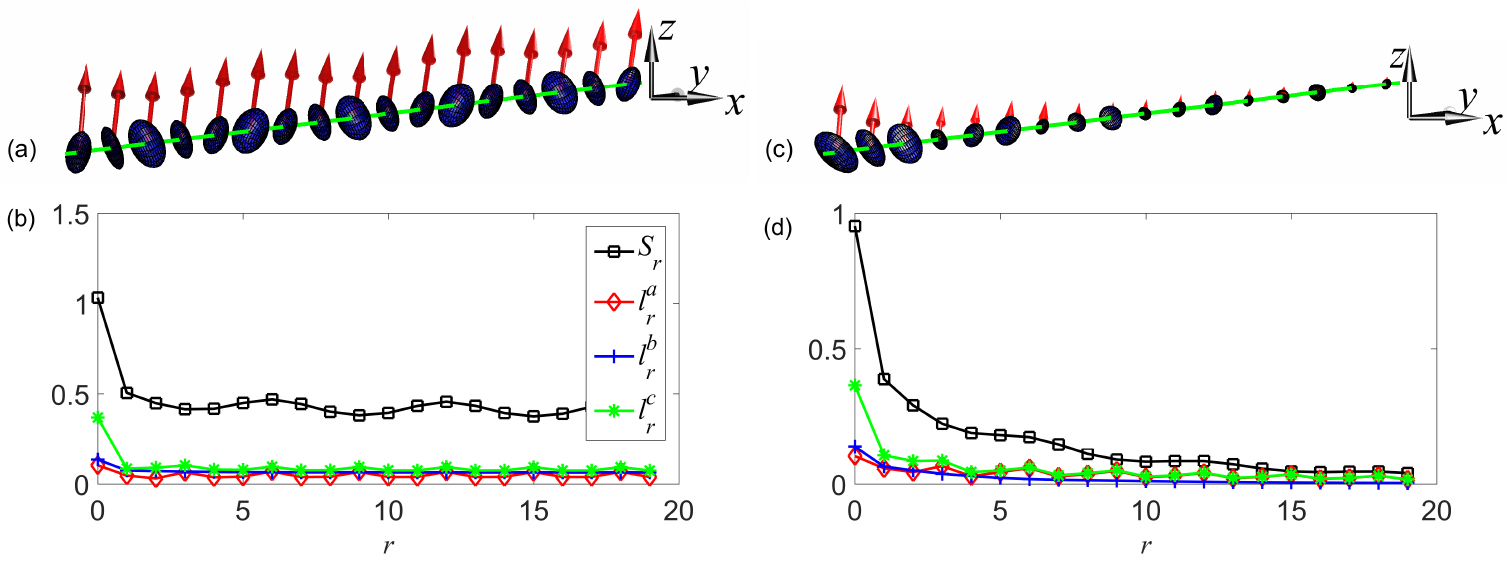} \hskip 0.0cm \centering
\caption{Spin-vector correlations arrows $\vec{\mathbb{S}}_{r}$ and
spin-tensor correlations ellipsoids $\mathbb{T}_{r}$ ($r=0,1,2,\cdots $).
(a)(b) FST phase with $\protect\theta /\protect\pi =0.9$ and $\Omega =0.4$.
(c)(d) Paramagnetic tensor phase with $\protect\theta /\protect\pi =0.9$ and
$\Omega =0.5$.}
\label{FST_P_corre}
\end{figure*}

Without the on-site Zeeman field ($\Omega =0$), the ferromagnetic spiral
tensor (FST) phase reduces to the ordinary ferromagnetic phase (F). The
corresponding local magnetic orders represented by the spin-vector density
arrows and spin-tensor density ellipsoids' axes are uniform in space, as
shown in Fig.~\ref{loop_FST}(a). For the FST with a small $\Omega $, the
spin-vector density arrows and spin-tensor density ellipsoids' axes
oscillate along the spin chain, forming spiral loops, as shown in Fig.~\ref%
{loop_FST}(b). With increasing $\Omega $, the spiral loops first enlarge
[see Figs.~\ref{loop_FST}(c)-\ref{loop_FST}(d)], then shrink to a narrow
ellipse [see Figs.~\ref{loop_FST}(e)-\ref{loop_FST}(g)]. Across the phase
transition point to the paramagnetic tensor phase (PT), the loops shrink to
lines, where the spin-vector density arrows become parallel to the Zeeman
field direction [see Fig.~\ref{loop_FST}(h)].

With increasing $\Omega $, the relative rotation between two neighboring
ellipsoids' axes becomes more significant [see Fig.~\ref{FST_PST_angle_length}(a) in the main text].
In the paramagnetic tensor phase, the relative rotation angle becomes $\pi
/2 $ [see Fig.~\ref{FST_PST_angle_length}(e) in the main text]. Both the size and orientation of the
ellipsoids oscillate along the spin chain, and a $\pi/2$ orientation change
is equivalent to a size deformation without rotation. Therefore in the
paramagnetic tensor phase, we only have the modulation in the ellipsoid size
[see Fig.~\ref{FST_PST_angle_length}(e) in the main text].

The ferromagnetic spiral order is characterized by the long-range
correlation of both spin-vector $\vec{\mathbb{S}}_{r}$ and spin-tensor $%
\mathbb{T}_{r}$. The spin arrow lengths $S_{r}$ and axis lengths of
ellipsoids $l_{r}^{n}$ exhibit power-low decay with the distance $r$, as
shown in Figs.~\ref{FST_P_corre}(a) and \ref{FST_P_corre}(b). In the
paramagnetic tensor phase, the vector correlation arrows $\vec{\mathbb{S}}%
_{r}$ and tensor correlation ellipsoids $\mathbb{T}_{r}$ decrease ($S_{r}$
and $l_{r}^{n}$ exhibit exponential decay) with the distance $r$, indicating
that there is no long-range order, as shown in Figs.~\ref{FST_P_corre}(c)
and \ref{FST_P_corre}(d).

\begin{figure*}[t]
\centering
\includegraphics[width = 18.0cm]{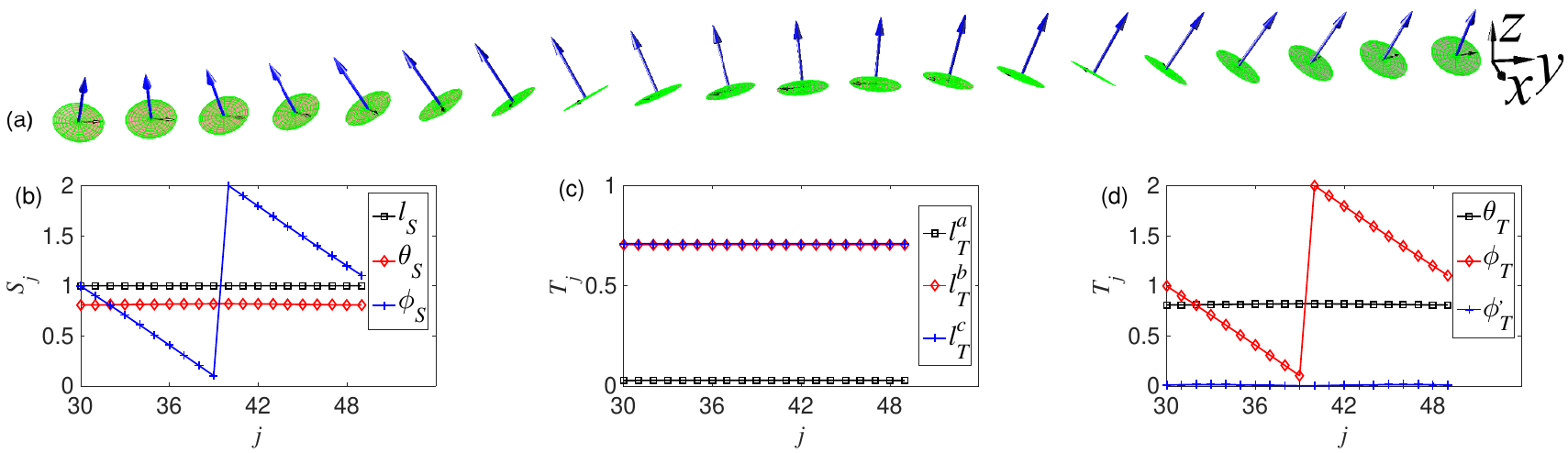} \hskip 0.0cm
\caption{Ferromagnetic spiral vector phase of Hamiltonian Eq. (14) in Ref. \cite{Pixley2017} with $\protect\theta /\protect\pi =1.1$, $%
\Omega =0.05$, $\protect\phi /\protect\pi =1/10$ and $L=80$. (a) Schematic
diagram of spin-vector density arrows $\vec{S_{j}}$ and spin-tensor density
ellipsoids $T_{j}$. (b) Spatial distributions of the spin-vector density
arrows $\vec{S_{j}}$. (c,d) Spatial distributions of the spin-tensor density
ellipsoids $T_{j}$. }
\label{SOC_SF_angle_length}
\end{figure*}

\section{\textbf{The results due to the spin-vector potential}}
\label{The results due to the spin-vector potential}

In contrast, in the ferromagnetic spiral vector phase induced by spiral
spin-vector Zeeman field in previous studies~\cite{Pixley2017} (tensor
results were not discussed in these works), the spin-vector density arrows
and spin-tensor density ellipsoids would rotate similarly with fixed
relative orientation [i.e. $\theta _{T}(j)-\theta _{S}(j)$, $\phi
_{T}(j)-\phi _{S}(j)$ and $\phi _{T}^{\prime }(j)$ are uniform along the
chain], and all ellipsoids have fixed sizes [i.e. $l_{T}^{n}(j)$ uniform
along the chain], as shown in Fig.~\ref{SOC_SF_angle_length}.

\section{\textbf{Dimer spiral tensor phase}}
\label{Dimer spiral tensor phase}

\begin{figure*}[t]
\centering
\includegraphics[width = 18.0cm]{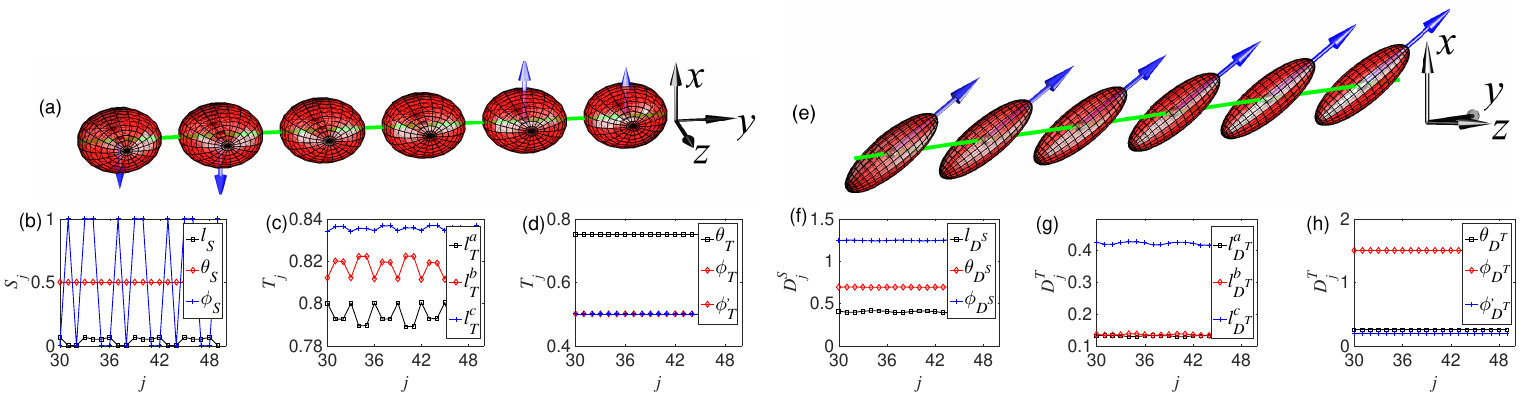} \hskip 0.0cm
\caption{The orders of the dimer spiral tensor phase.
(a) Schematic diagram of spin-vector density
arrows $\vec{S_{j}}$
and spin-tensor density ellipsoids $T_{j}$. (b)
Spatial distributions of
the spin-vector density arrows $\vec{S_{j}}$.
(c,d) Spatial
distributions of the spin-tensor density ellipsoids
$T_{j}$.
(e) Schematic diagram of dimer-vector density arrows
$D^{\vec{S}}_{j}$
and dimer-tensor density ellipsoids $D^T_{j}$. (f)
Spatial distributions of
the dimer-vector density arrows $D^{\vec{S}}_{j}$.
(g,h) Spatial
distributions of the dimer-tensor density ellipsoids
$D^T_{j}$. In all subfigures, $\protect\theta /\protect\pi =1.4$, $\Omega
=0.1$, $\protect\phi /\protect\pi =1/6$, and $L=96$.}
\label{DST_site}
\end{figure*}

\begin{figure*}[t]
\centering
\includegraphics[width = 15cm]{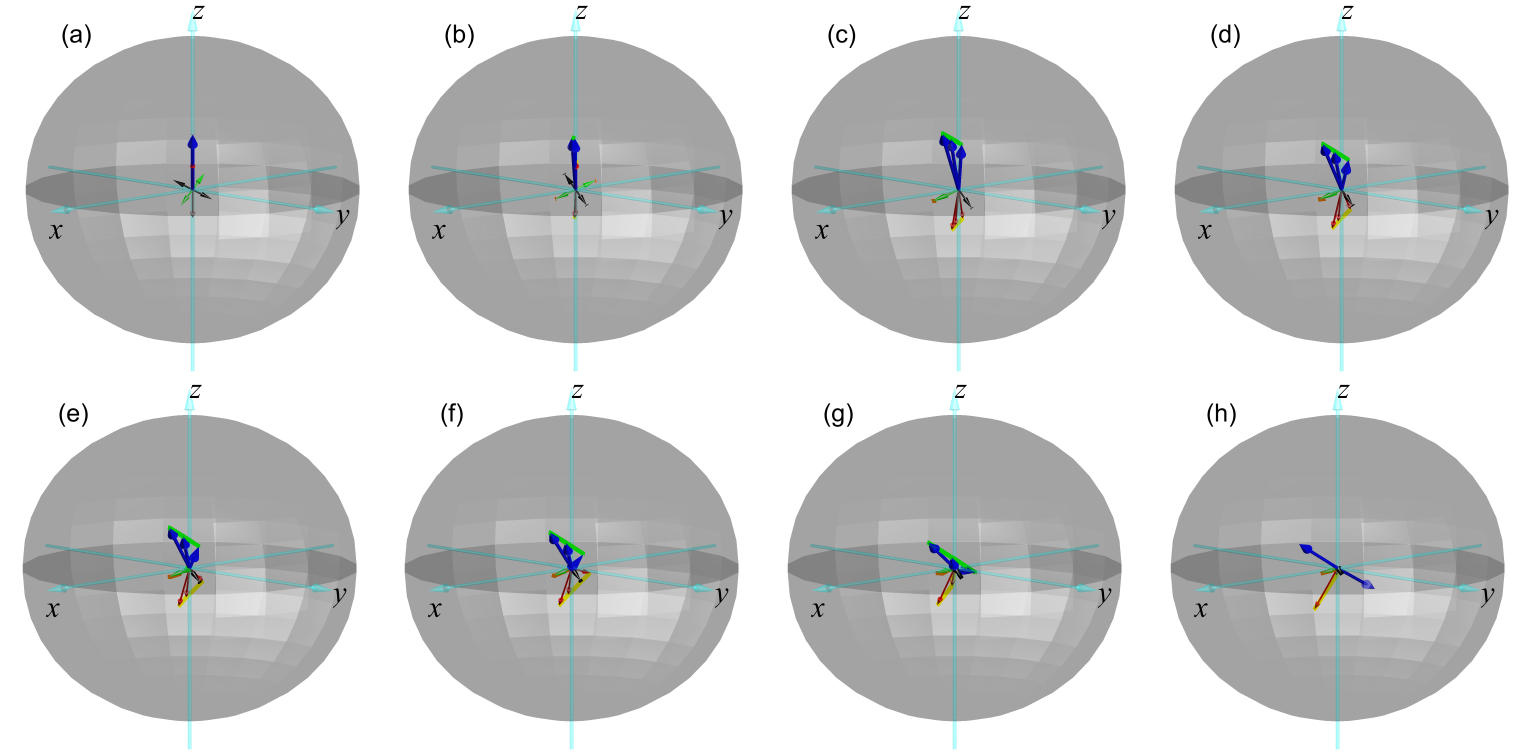} \hskip 0.0cm
\caption{Phase transition between D-DST-P. Blue arrows represent
spin-vectors and green line represents the corresponding loop for the spiral
loop of the dimer-vector density arrows $D_{j}^{\vec{S}}$. Red, green and
black arrows represent three axes of ellipsoids, and yellow, orange, and
black lines represent corresponding loops for dimer-tensor density
ellipsoids $D_{j}^{T}$. (a) $\Omega =0.0$, (b) $\Omega =0.1$, (c) $\Omega
=0.18$, (d) $\Omega =0.2$, (e) $\Omega =0.21$, (f) $\Omega =0.22$, (g) $%
\Omega =0.25$, (h) $\Omega =0.3$. The phase transition occurs at $\Omega
_{c}=0.25$. In all subfigures, $\protect\theta /\protect\pi =1.4$, $\protect%
\phi /\protect\pi =1/6$ and $L=96$.}
\label{loop_DST}
\end{figure*}

\begin{figure*}[t]
\centering
\includegraphics[width = 8cm]{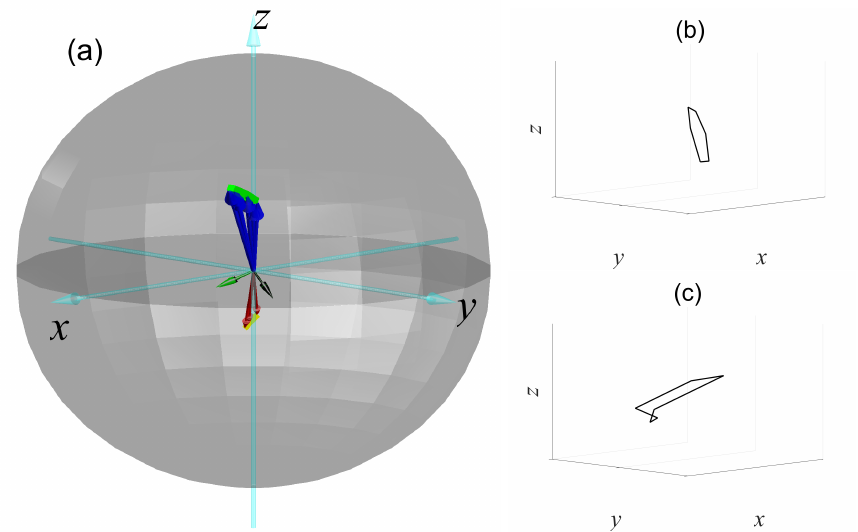} \hskip 0.0cm
\caption{Spiral loops for the DST with a phase shift $\protect\pi /15$
between vector and tensor terms in the on-site Zeeman field modulation along
the chain. (a) Blue arrows represent spin-vectors, and green line represents
the corresponding loop for the spiral loop of dimer-vector density arrows $%
\mathrm{D}^{\vec{S}}_{j}$. Red, green and black arrows represent three axes
of ellipsoids, while yellow, orange and black lines represent corresponding
loop for the dimer-tensor densities ellipsoids $\mathrm{D}^{T}_{j}$. $\Omega
=0.18$, $\protect\theta /\protect\pi =1.4$, $\protect\phi /\protect\pi =1/6$
and $L=96$. (b) The enlarged loop of dimer-vector density arrows. (c) The
enlarged loop of one axe (red arrows) of dimer-tensor density ellipsoids.}
\label{DST_018_shift}
\end{figure*}

\begin{figure*}[t]
\centering
\includegraphics[width = 17cm]{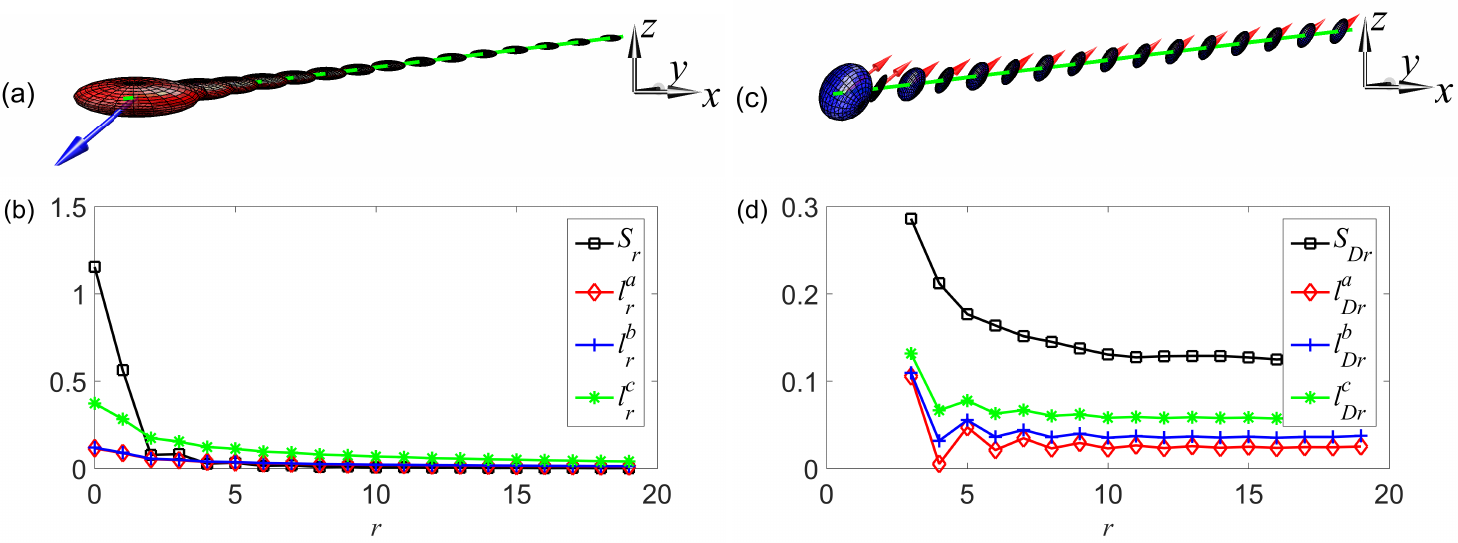} \hskip 0.0cm
\caption{(a)(b) Spin-vector correlation arrows $\vec{\mathbb{S}}_{r}$ and
spin-tensor correlation ellipsoids $\mathbb{T}_{r}$. (c)(d) Dimer-vector
correlations arrows $\mathbb{D}^{\vec{\mathbb{S}}}_{r}$ and dimer-tensor
correlations ellipsoids $\mathbb{D}^{T}_{r}$. In all subfigures, $\protect%
\theta /\protect\pi =1.4$, $\Omega =0.1$, $\protect\phi /\protect\pi =1/6$,
and $L=96$.}
\label{DST_corre}
\end{figure*}

The spin-1 Hamiltonian may support dimer orders in certain parameter region
(i.e. $1.25<\theta /\pi \leq 1.5$), which describe the pairing order between
two neighboring sites. The corresponding dimer-vector and dimer-tensor
operators are defined as $\mathrm{D}^{S^{\alpha}}_{j}=(-1)^{j}\langle
S_{j-1}^{\alpha }S_{j}^{\alpha }-S_{j}^{\alpha }S_{j+1}^{\alpha }\rangle $
and $\mathrm{D}^{T^{\alpha \beta}}_{j}=(-1)^{j}\langle N_{j-1}^{\alpha
\beta }N_{j}^{\alpha \beta }-N_{j}^{\alpha \beta }N_{j+1}^{\alpha \beta
}\rangle $. For instance, the dimer-vector $\mathrm{D}^{S^{\alpha}}_{j}$
and dimer-tensor $\mathrm{D}^{T^{\alpha \beta}}_{j}$ are nonzero for the
spin singlet pairing between neighboring sites in the dimer phase for the
Hamiltonian $H_{\mathrm{spin}}^{\mathrm{0}}$ even without the spiral Zeeman
field. Such dimer-vector density $\mathrm{D}^{\vec{S}}_{j}$ and dimer-tensor
density $\mathrm{D}^{T}_{j}$ can also be described geometrically using
arrows and ellipsoids, similar as the spin-vector and spin-tensor for a
single site.
There are also 8 dimer-moments
(i.e., 3 dimer-vectors and 5 dimer-tensors: the length $l_{D^S}$ and spherical coordinates $\theta _{D^S}$, $\phi _{D^S}$ of the arrow, the two axis lengths $l_{D^T}^{a,b}$ with the third
axis length $l_{D^T}^{c}= \sqrt{{2-(l_{D^S})}^2- {(l_{D^T}^{a})}^2 -{(l_{D^T}^{b})}^2}$
and orientational Euler angles $\theta_{D^T}$, $\phi_{D^T}$, $\phi
_{D^T}^{\prime}$ of the ellipsoid).
Similarly, dimer-vector correlation $\mathbb{D}^{\vec{S}}_{r}$
with elements $\mathbb{D}^{S^{\alpha}}_{r}=\mathbb{C}(\mathrm{D}^{{S}^{\alpha}},r)$ can be described by arrows with $S_{Dr}$ the length of
arrows, and dimer-tensor correlation $\mathbb{D}^{T}_{r}$ can be described by ellipsoids with principle axis
lengths $l_{\mathbb{D}r}^{n}$ ($n=a,b,c$) given by the matrix $\mathbb{D}^{{T}^{\alpha \beta}}_{r}=\mathbb{C}(\mathrm{D}^{{T}^{\alpha \beta
}},r)$, respectively.

The dimer spiral tensor phase (DST) is non-degenerate, therefore the
behaviors of spin-vector arrows and spin-tensor ellipsoids are the same as those in the
paramagnetic tensor phase [see Fig.~\ref{DST_site}(a)].
Only the spin-tensor density ellipsoid's lengths $l_{T}^{n}\left( j\right) $ oscillate, while the angles $\theta _{T}(j)$, $\phi _{T}(j)$
and $\phi_{T}^{\prime}(j)$ are
uniform. Specifically, the size of spin-vector density arrows $\vec{S_{j}}$
and spin-tensor density ellipsoids $T_{j}$ oscillate along the chain,
forming lines in the Bloch sphere that are similar as those for the
paramagnetic tensor phase [see Figs.~\ref{DST_site}(b)-\ref{DST_site}(d)].
However, the dimer-vector density $\mathrm{D}^{\vec{S}}_{j}$ and dimer-tensor
density $\mathrm{D}^{T}_{j}$ can form spiral loops for a finite $\Omega $,
as shown in Fig.~\ref{DST_site}(e).
The $l_{D^S}(j)$, $\theta_{D^S}(j)$ and $\phi_{D^S}(j)$ oscillate along
the chain [see Fig.~\ref{DST_site}(f)]. The dimer-tensor ellipsoids' lengths $l_{\mathrm{D}^T}^{n}(j)$ and Euler angles $\theta_{D^T}(j)$, $\phi_{D^T}(j)$, $\phi
_{D^T}^{\prime}(j)$ oscillate along
the chain [$\theta_{D^T}(j)$, $\phi_{D^T}(j)$ oscillate differently from $\theta_{D^S}(j)$, $\phi_{D^S}(j)$] [see Figs.~\ref{DST_site}(g) and \ref{DST_site}(h)], originating from
different spiral loops for the dimer-vector arrows $\mathrm{D}^{\vec{S}}_{j}$ and
the dimer-tensor ellipsoid $\mathrm{D}^{T}_{j}$.

For the ordinary dimer phase (D) with $\Omega =0$, the dimer-vector density
arrows and dimer-tensor density ellipsoids' axes are uniform in space, as
shown in Fig.~\ref{loop_DST}(a). In the dimer spiral tensor phase with
increasing $\Omega $, the spiral loop for the dimer order first enlarges
[see Figs.~\ref{loop_DST}(b)-\ref{loop_DST}(g)], then shrinks into a point
at the phase transition to the paramagnetic tensor phase [see Fig.~\ref%
{loop_DST}(h)]. For the on-site Zeeman field given in the main text, the
system has a mirror symmetry, therefore the spiral loops formed by the dimer
densities shrink to lines. If a phase shift between spin-vector and -tensor
terms is introduced to the on-site Zeeman field modulation along the chain,
the mirror symmetry is broken and the dimer spiral loops would emerge as
circles. In addition, the dimer-vector density $\mathrm{D}^{\vec{S}}_{j}$
and dimer-tensor density $\mathrm{D}^{T}_{j}$ form different spiral loops,
leading to relative rotations between them, as shown in Fig.~\ref%
{DST_018_shift}.

There is no long-range ordinary correlations of spin-vector $\vec{\mathbb{S}}%
_{r}$ and spin-tensor $\mathbb{T}_{r}$ in the dimer spiral tensor phase [see
Figs.~\ref{DST_corre}(a) and \ref{DST_corre}(b)], which are the same as the
paramagnetic tensor phase [see Figs.~\ref{FST_P_corre}(c) and \ref%
{FST_P_corre}(d)]. Instead, the system possesses long-range dimer
correlations for both dimer-vector $\mathbb{D}^{\vec{\mathbb{S}}}_{r}$ and
dimer-tensor $\mathbb{D}^{T}_{r}$, the length of arrows $S_{Dr}$ and the
axes length of ellipsoids $l_{Dr}^{n}$ ($n=a,b,c$) exhibit power-low decay
and have long-range correlations [see Figs.~\ref{DST_corre}(c) and \ref%
{DST_corre}(d)].

\section{\textbf{Correlation lengths near phase transitions}}
\label{Correlation lengths near phase transitions}

\begin{figure*}[t]
\centering
\includegraphics[width = 12cm]{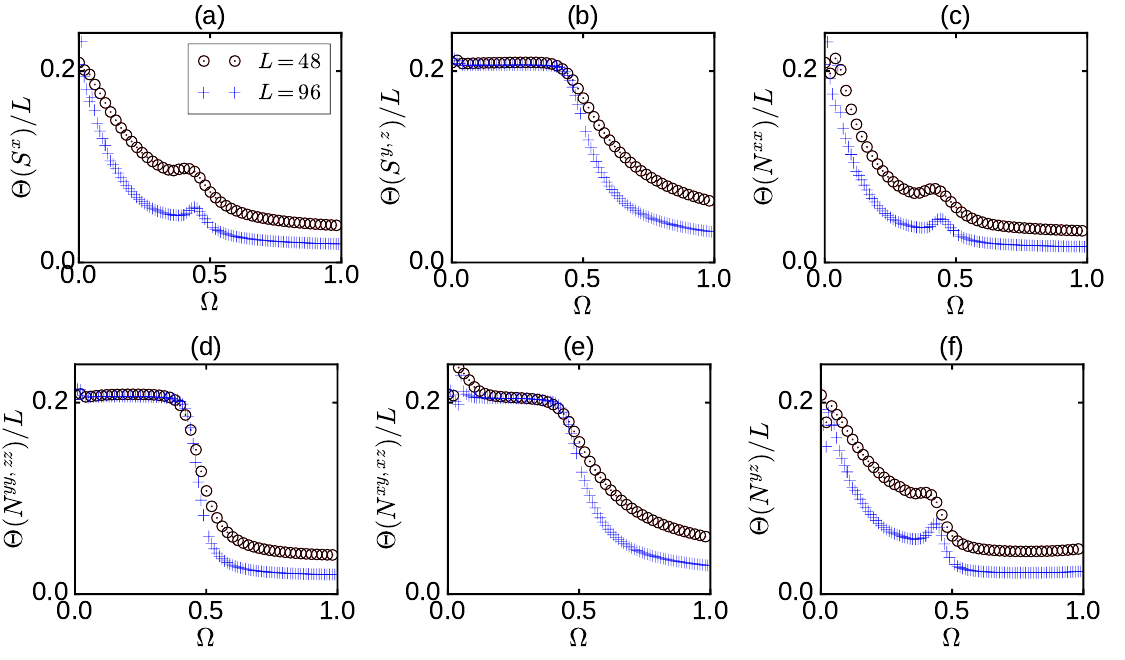} \hskip 0.0cm
\caption{Spin-vector(-tensor) correlation lengths $\Theta (\hat{O})$ as
functions of $\Omega $ for different lattice length $L$ across the phase
transition between FST and P. In all subfigures, $\protect\theta /\protect%
\pi =0.9$ and $\protect\phi /\protect\pi =1/6$.}
\label{corre_length_SFST}
\end{figure*}

\begin{figure*}[t]
\centering
\includegraphics[width = 12cm]{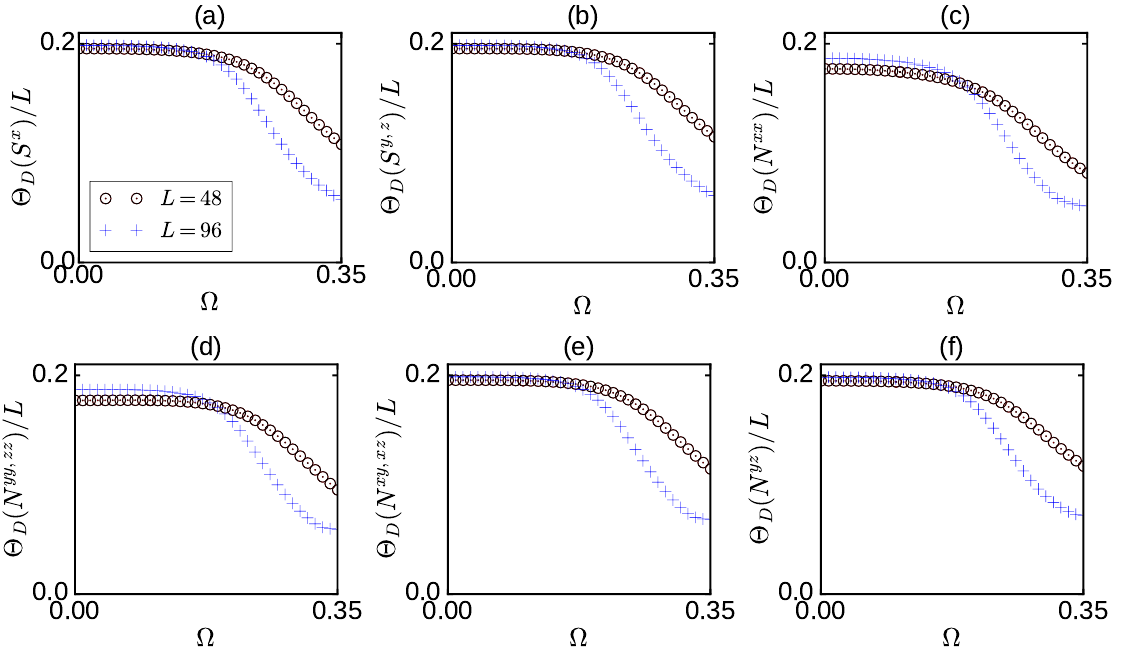} \hskip 0.0cm
\caption{Dimer-vector(-tensor) correlation lengths $\Theta _{\mathrm{D}}(%
\hat{O})/L$ as functions of $\Omega $ for different lattice length $L$
across the phase transition between DST and P. In all subfigures, $\protect%
\theta /\protect\pi =1.4$ and $\protect\phi /\protect\pi =1/6$.}
\label{corre_length_Dimer_SDST}
\end{figure*}

The critical point $\Omega _{c}$ for the phase transition between
ferromagnetic spiral tensor and paramagnetic tensor phases in the
thermodynamic limit can be examined by the spin-vector(-tensor) correlation
lengths $\Theta (\hat{O})$ ($\hat{O}=S^{y,z},T^{yy,zz,xy,xz}$). The DMRG
results show that $\Theta (\hat{O})/L$ for several finite lattice lengths
cross at one point with increasing $\Omega $, which is the critical point $%
\Omega _{c}$ between ferromagnetic spiral tensor and paramagnetic tensor
phases, as shown in Fig.~\ref{corre_length_SFST}.

Similarly, the phase transition between dimer spiral tensor and paramagnetic
tensor phases in the thermodynamic limit can be examined by the
dimer-vector(-tensor) correlation lengths $\Theta _{\mathrm{D}}(\hat{O})/L$ (%
$\hat{O}={S^{x,y,z},T^{xx,yy,zz,xy,xz,yz}}$). There is (no) long-range dimer
correlation in the thermodynamic limit for the dimer spiral tensor
(paramagnetic tensor) phase. With increasing $\Omega $, the dimer
correlation length $\Theta _{\mathrm{D}}(\hat{O})/L$ for a finite lattice
retains and then decays. In the decay region, $\Theta _{\mathrm{D}}(\hat{O}%
)/L$ for several finite lattice lengths cross at one point, which
corresponds to the phase transition critical point $\Omega _{c}$, as shown
in Fig.~\ref{corre_length_Dimer_SDST}.

\section{\textbf{Magnetic order with spin-tensor interaction}}
\label{Magnetic order with spin-tensor interaction}

\begin{figure}[t]
\centering
\includegraphics[width = 6.0cm]{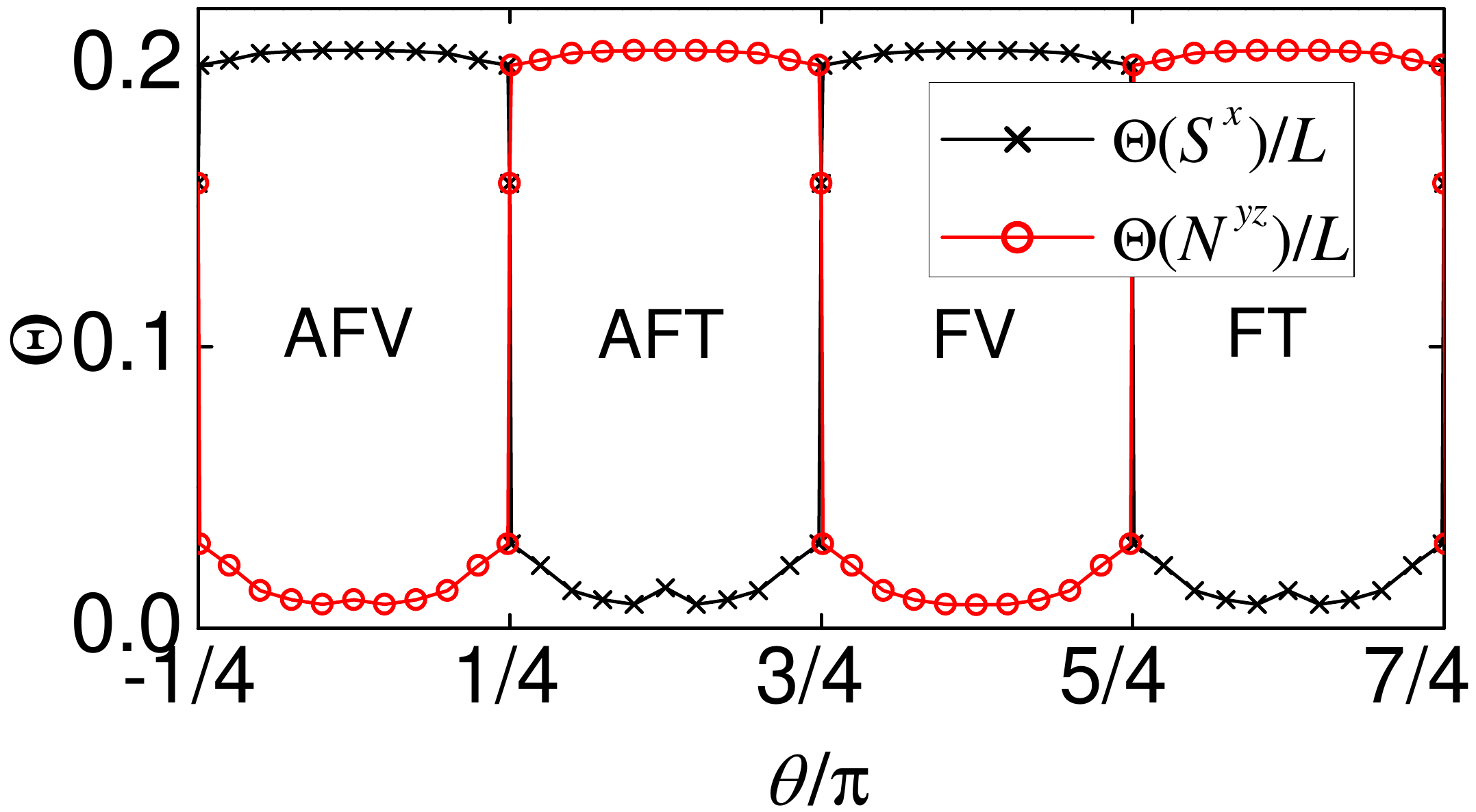} \hskip 0.5cm
\caption{Phase diagrams, the spin-vector correlation length $\Theta
(S^{x})/L $, and the spin-tensor correlation length $\Theta (N^{yz})/L$ as
functions of $\protect\theta $. The phases are the ferromagnetic vector
(FV), antiferromagnetic vector (AFV), ferromagnetic tensor (FT), and
antiferromagnetic tensor order (AFT). $\Omega =0.1$, $\protect\phi /\protect%
\pi =1/6$ and $L=96$.}
\label{phase diagram_corre}
\end{figure}

The spin-tensor can appear not only as an on-site Zeeman field, but also as
interactions between nearest-neighbor sites. The biquadratic term of
Hamiltonian~(\ref{HS}) in the main text contains many types of spin-tensor
interactions, making it hard to identify the spin-tensor correlations
induced by each term. Moreover, the spin-tensor correlations cannot be
isolated out because they are bound with the spin-vector correlations. Here
we consider a simple toy spin Hamiltonian,
\begin{eqnarray}
H_{\mathrm{spin}}\!\! &=& \!\! \sum_{j}J_{\mathrm{a}}S_{j}^{x}S_{j+1}^{x}+4J_{\mathrm{b}%
}N_{j}^{yz}N_{j+1}^{yz}+2\Omega \cos \left( 2\phi j\right) S_{j}^{x} \notag \\
&-& 2\Omega
\sin \left( 2\phi j\right) S_{j}^{y},  \label{HStso0}
\end{eqnarray}
where $J_{\mathrm{a}}=\cos \theta $ and $J_{\mathrm{b}}=\sin \theta $. The
competition between the spin-vector and spin-tensor interactions induces
ferromagnetic or antiferromagnetic vector or tensor phases for different $%
\theta $, as shown in Fig.~\ref{phase diagram_corre}.

\begin{figure}[t]
\centering
\includegraphics[width = 8.0cm]{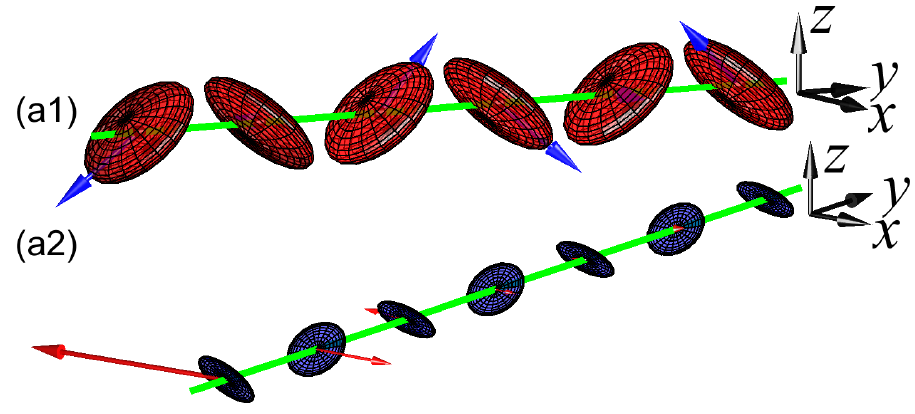} \hskip 0.5cm
\caption{ Spin-vector and -tensor order and correlation in the AFT. (a1)
Spin-vector density arrows $\vec{S_{j}}$ and spin-tensors density ellipsoids
$T_{j}$. (a2) Spin-vector correlation arrows $\vec{\mathbb{S}}_{r}$ and
spin-tensors correlation ellipsoids $\mathbb{T}_{r}$ ($r=1,2,\cdots $). In
all subfigures, $\protect\theta /\protect\pi =0.3$, $\Omega =0.1$, $\protect%
\phi /\protect\pi =1/6$ and $L=96$. The arrows are enlarged by 8 times.}
\label{interaction_Vector_Tensor}
\end{figure}

A weak spiral zeeman field (here we take $\Omega =0.1$) can induce spiral
spin-vector densities, but does not affect the long-range magnetic order. We
find that the ferromagnetic (antiferromagnetic) vector phases are similar as
those discussed in Ref.~\cite{Pixley2017}, where the system possesses
both long-range vector and tensor correlations with spiral spin-vectors
densities, and the long-range tensor correlations are induced by the
spin-vector interactions. In ferromagnetic (antiferromagnetic) tensor phase
[FT (AFT)], the system only possesses long-range spin-tensor correlations
(with no long-range spin-vector correlations), which are induced directly by
the spin-tensor interactions.
In the AFT phase, the spin-tensor density ellipsoids $T_{j}$ have orthogonal
axes $\diagdown \diagup \diagdown \diagup \diagdown \diagup $ between
nearest-neighbor sites,
while in the FT phase, the axes of the ellipsoids $T_{j}$ are parallel $%
\diagdown \diagdown \diagdown \diagdown $ between nearest-neighbor sites. In
both the FT and AFT phases, the spin-vector density arrows $\vec{S}_{j}$ are
spiral along the chain. Such FT and AFT phases due to the spin-tensor
interactions are very different from previous studies of spin-vector
interactions. As examples, in Figs.~\ref{interaction_Vector_Tensor}(a1)-\ref%
{interaction_Vector_Tensor}(a2), we show the magnetic local densities and
non-local correlations in the AFT phase.

\end{document}